\pdfoutput=1
\documentclass[sigconf,nonacm]{acmart}

\usepackage{tabularx,array,multirow}
\usepackage[ruled,vlined,linesnumbered]{algorithm2e}
\usepackage{enumitem,xcolor}
\usepackage{tikz}
\usepackage{etoolbox}

\SetAlFnt{\small}
\SetAlCapFnt{\small}
\SetAlCapNameFnt{\small}
\SetAlgoNlRelativeSize{0}

\newcommand{\design}{\textsc{Qlash}}

\newcommand{\circledtext}[1]{%
  \tikz[baseline=(n.base)]{%
    \node[draw, circle, line width=0.45pt, inner sep=0pt,
          minimum size=1.2em, font=\normalfont\normalsize\bfseries,
          text height=0.68em, text depth=0pt] (n) {#1};%
  }%
}

\setcopyright{none}

\makeatletter
\patchcmd{\@mktitle@iii}{\par\bigskip}{\par\bigskip\vskip-3pt}{}{%
  \PackageWarningNoLine{qlash}{Could not adjust title-to-author spacing}%
}
\makeatother

\AtBeginDocument{%
  \hypersetup{colorlinks=true,allcolors=blue}%
}

\begin{document}

\title{\textit{The \textsc{Qlash} --- Should I Stay or Should I Grow?}\\
  Syndrome-Adaptive Logical Bell-Pair Preparation}

\author{%
  Won Joon Yun\texorpdfstring{\textsuperscript{\textdagger}}{},\quad
  Maxwell Poster\texorpdfstring{\textsuperscript{*}}{},\quad
  Sayam Sethi\texorpdfstring{\textsuperscript{*}}{},\quad
  Jonathan Baker%
}
\renewcommand{\shortauthors}{Yun et al.}

\affiliation[obeypunctuation=true]{%
  \institution{The University of Texas at Austin}%
  \city{}%
  \state{}%
  \country{}%
}

\begin{abstract}
Distributed quantum computing requires Logical Bell Pairs (LBPs) for remote operations.
Preparing them, however, consumes Physical Bell Pairs (PBPs) in costly repeat-until-success processes that occupy physical qubits across repeated syndrome measurements, incurring significant spacetime costs.
Entanglement Boosting (EB) uses the complementary gap, a measure of decoding confidence, for postselection after a fixed number of initial rounds.
These extra rounds consume spacetime, which may introduce errors that lead to more discards, and require additional PBPs for retries.
Moreover, LBP preparation under EB starts at the delivered code distance.
This is problematic as distance-5 rounds use $2.9\times$ the spacetime of a distance-3 round, regardless of policy or postselection.

We propose \textsc{Qlash}, a compiler pass that uses intermediate complementary gaps to choose whether to remeasure, grow, discard, or deliver an LBP at the required distance.
This reduces the spacetime requirement per LBP at the required code distance while keeping the infidelity of delivered LBPs below the target.
\textit{Offline, }we collect circuit-level simulation data to model how remeasurement and growth affect the complementary gap and LBP infidelity.
Using this model, we optimize a policy that jointly selects the starting code configuration and subsequent actions to reduce PBP consumption and spacetime per delivered LBP within the infidelity target.
This policy  is stored as a lookup table.
\textit{Online,} after each syndrome-measurement round, the controller uses the observed complementary gap to read the corresponding table entry.
Compared with EB, \textsc{Qlash} avoids unnecessary syndrome-measurement rounds and discards LBPs at smaller code distances before incurring the cost at the required distance.
Using circuit-level simulations at matched delivered-LBP
infidelity, with a $2\times10^{-5}$ target infidelity and PBP infidelities of 2.5\%, 5\%, and 6\%, \textsc{Qlash} uses 18.8--25.2\% fewer PBPs and 19.6--68.9\% less spacetime per delivered LBP than EB.
\end{abstract}

\maketitle

\begingroup
  \renewcommand{\thefootnote}{\fnsymbol{footnote}}
  \footnotetext[1]{Maxwell Poster and Sayam Sethi contributed equally to this work.}
  \footnotetext[2]{Won Joon Yun:
    \href{mailto:wonjoon.yun@utexas.edu}{\nolinkurl{wonjoon.yun@utexas.edu}}.}
\endgroup

\section{Introduction}
\begin{figure}[t]
    \centering
    \includegraphics[width=\columnwidth]{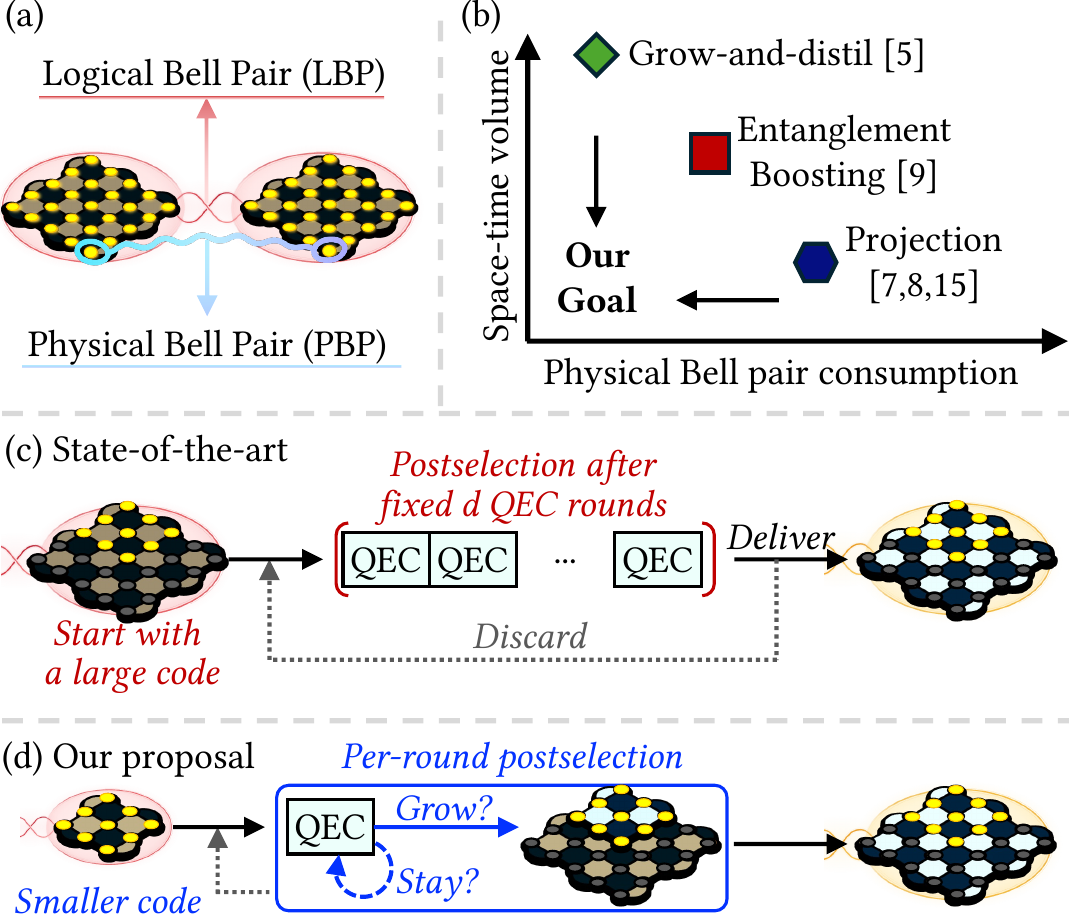}
    \vspace{-20pt}
    \caption{(a) Illustration of LBP. (b) Goal: reduce PBP consumption and spacetime volume. (c) EB starts at the target distance and postselects after a fixed number of QEC rounds. (d) Our per-round syndrome-guided postselection. Yellow and dark gray dots show PBPs and data qubits, respectively.}
    \label{fig:1}
    \vspace{-15pt}
\end{figure}

Distributed quantum computers connect chips to support computations beyond the qubit capacity of a single chip~\cite{duan2010,jiang2007,nickerson2013,nickerson2014,marqversen2025}. Fault-tolerant operations across chips use Logical Bell Pairs (LBPs), each consisting of multiple Physical Bell Pairs (PBPs) on different chips~\cite{marqversen2025,pattison2025,bonillaataides2025,maeda2025,sunami2025,bennett1993,eisert2000}, as shown in Figure~\ref{fig:1}(a).
For a reliable remote operation, LBPs must be distilled, which requires repeated syndrome measurements. These measurements occupy qubits over time, incurring spacetime volume overheads. A limited PBP supply rate can become a major latency bottleneck in LBP preparation.
Each preparation requires multiple PBPs and must wait if too few are available. For example, generating one PBP in a trapped-ion DQC demonstration took an average of $103\,\mathrm{ms}$, approximately $200\times$ as long as a local mid-circuit measurement ($500\,\mu\mathrm{s}$)~\cite{main2025}.
Thus, as illustrated in Figure~\ref{fig:1}(b), the goal is to reduce both PBP consumption and spacetime per delivered LBP while meeting the required code distance and target infidelity.

\begin{figure*}
\centering
\vspace{-10pt}
\includegraphics[width=\textwidth]{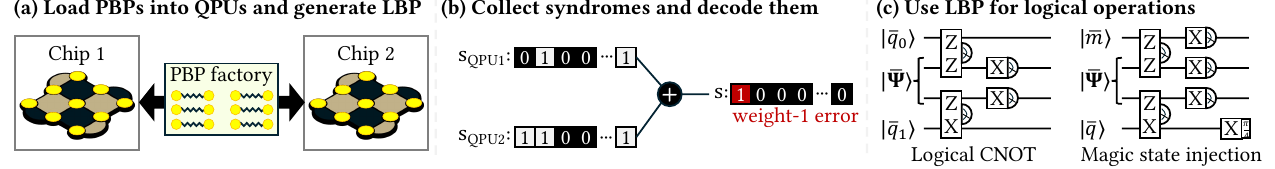}
\vspace{-22pt}
    \caption{LBP preparation and use across two chips: (1) PBP encoding, (2) joint-syndrome decoding, and (3) LBP usages.}
\label{fig:2}
\vspace{-15pt}
\end{figure*}

Prior works on LBP preparation improve fidelity through Quantum Error Correction (QEC) or error detection~\cite{bennett1996mixed,glancy2006,pattison2025,marqversen2025,sunami2025}.
\textit{Direct code projection} initializes all data qubits of two code blocks with PBP halves and measures stabilizers to project the pairs into the code space.
Measurement outcomes from both chips form a joint syndrome for error correction~\cite{bonillaataides2025,maeda2025}.
Without postselection, every preparation is retained. At a fixed code distance, however, increasing the PBP error rate can prevent this approach from meeting the target infidelity~\cite{bonillaataides2025}.
\textit{Projection with quantum error detection} can further reduce delivered-LBP infidelity by accepting only preparations with no detected errors in the joint syndrome~\cite{shi2025stabilizer}. However, this rule also discards LBPs with correctable errors, requiring additional preparations and increasing PBP consumption and spacetime per delivered LBP. \textit{Entanglement Boosting (EB)}, the state-of-the-art LBP preparation method, further reduces \textit{both} PBP consumption and spacetime overheads~\cite{sunami2025}. It reduces the initial PBP count by initializing only a small region of each code block with PBP halves and preparing the remaining data qubits locally, as shown in Figure~\ref{fig:1}(c). EB also uses the \textit{complementary gap}, a measure of confidence in the correction selected by the decoder, to retain some LBPs with detected errors when their gaps are sufficiently large. This postselection occurs after a fixed number of QEC rounds.

However, the state-of-the-art has two drawbacks. \textit{First}, every postselection decision is made after the $d$ QEC rounds before postselection, regardless of the intermediate syndromes. Additional rounds provide more syndrome information, but syndrome measurement itself may introduce errors. In our circuit-level simulation for distance-5 LBP preparation, the conditional failure rate among the lowest-gap outputs is over 10$\times$ higher after five rounds than after just one. Subsequently, this failure rate dictates the overall LBP infidelity. \textit{Second}, EB uses the target code distance (typically a large code block) from the first QEC round, so attempts that are later discarded incur the substantial cost of operating at that distance. This is because a distance-$d$ surface-code block requires $\mathcal{O}(d^2)$ physical qubits, increasing the spacetime cost of each QEC round at larger distances~\cite{horsman2012}. Thus, preparing the LBP directly at distance 5 results in nearly $3\times$ more discarded physical qubits than first preparing it at distance 3 and then conditionally growing it to distance 5. These findings motivate a fine-grained postselection protocol that uses information from each QEC round to decide whether to continue preparation at the current code distance, grow the code, or stop.

We propose \design{}, which selects whether to remeasure, grow, deliver, or discard every QEC round using its complementary gap, as shown in Figure~\ref{fig:1}(d).
This is not trivial, because the per-round gap from each QEC round alone does not predict how remeasurement or growth will affect LBP infidelity and the resources required per delivered LBP. It is also unclear whether starting at a smaller code distance minimizes the resources required per delivered LBP. Lastly, each decision must be made quickly to prevent extending QEC cycle time.
Thus, \design{} employs a two-stage approach. \textit{Offline,} we formulate LBP preparation as a Markov decision process. We collect circuit-level simulation data to estimate how remeasurement and code growth affect the complementary gap and LBP infidelity. For each current state and action, we estimate transition probabilities by using the observed transition counts over all next states. For a given preparation policy, the model estimates how many further rounds and retries are needed on average before delivery. Linear programming then jointly selects the starting code and per-round decisions to minimize the PBP and spacetime cost per delivered LBP under the infidelity limit.
\textit{Online,} \design{} uses the optimized LBP policy as a look-up table, so that it makes a decision quickly.

Distance-7 LBP preparation shows 18.8--25.2\% lower PBP use and 19.6--68.9\% lower active spacetime per delivery than EB at matched delivered-LBP infidelity. Once the complementary gap is available, the distance-7 FPGA controller for \design{} selects an action in~25\,ns.

\vspace{-0.01in}
This paper makes the following contributions:\\[-12pt]
\begin{itemize}[leftmargin=1.1em,labelsep=0.3em,nosep]
    \item We show that LBP infidelity depends on both the QEC round count and the per-round complementary gap, and that starting with a smaller LBP reduces preparation overhead.\\[-12pt]
    \item We propose \design{} that selects which codes to start with and whether to remeasure, grow, discard, or deliver every QEC round.\\[-12pt]
    \item We implement a lightweight FPGA controller for the compiled \design{}, which takes only 25\,ns for a postselection decision.
\end{itemize}

\newpage

\section{Background}
\subsection{LBPs in Fault-Tolerant DQCs: An Overview}

A Distributed Quantum Computer (DQC) is an architecture that connects Quantum Processing Units (QPUs) through photonic interconnects to support large-scale quantum computation~\cite{duan2010,jiang2007,nickerson2013,nickerson2014,marqversen2025}. In fault-tolerant DQCs, logical operations are executed either locally within a QPU or across multiple QPUs, depending on where the participating logical qubits reside. If they are within the same QPU, their operation is implemented fault-tolerantly via transversal operations or lattice surgery~\cite{gottesman1998,gottesman1999,horsman2012,fowler2012,litinski2019}. If the logical qubits are located on different QPUs, remote operations can be implemented through teleportation using a shared quantum channel between QPUs and classical communication \cite{pattison2025,bonillaataides2025,maeda2025,marqversen2025,sunami2025, bennett1993,eisert2000}.

This shared channel is implemented through an LBP, which consists of two logical qubits prepared in a Bell state~\cite{pattison2025,bonillaataides2025,maeda2025,marqversen2025,sunami2025}. To generate an LBP, the two QPUs use PBPs generated over their photonic interconnects, as illustrated in Figure~\ref{fig:2}(a). Each QPU places its half of each PBP at the position of a data qubit of a code block. The QPUs then perform a syndrome measurement circuit to make parity checks detect error events~\cite{maeda2025,sunami2025}. Since each LBP half is logically encoded using the QEC code~\cite{pattison2025,bonillaataides2025,maeda2025,marqversen2025,sunami2025}, the shared channel can be maintained through QEC until the LBP is consumed. Since each LBP half is correlated with the other LBP half, their \textit{joint parity checks} with the bit-wise XOR are used as illustrated in Figure~\ref{fig:2}(b).

Once the LBP is ready to use, the remote operation can employ gate or state teleportation by consuming the LBP~\cite{gottesman1999,eisert2000,bennett1993,duan2010,marqversen2025}. Figure~\ref{fig:2}(c) illustrates two representative remote operations in the fault-tolerant DQCs using the LBP. Each logical qubit performs a Pauli-product measurement with an LBP half. Then, the LBP is measured. Their measured outcomes are transferred to the other QPU for the Pauli correction of their logical qubits~\cite{eisert2000,horsman2012,litinski2019}. For magic state injection into a logical qubit on a different QPU, an LBP is consumed for teleportation~\cite{bennett1993,gottesman1999}.

\subsection{Complementary Gap for LBP Postselection}
\label{sec:2.2}
The PBP error rate ($p_{\mathrm{Bell}}$) is approximately $3$--$6\%$~\cite{main2025,stephenson2020}, more than an order of magnitude above local CNOT infidelities of  $p\simeq10^{-3}$~\cite{bluvstein2026architecture}. An LBP is therefore more vulnerable to logical failure than a data block. Consequently, remote operations become more error-prone than local operations. To prevent this, postselection of LBPs is adopted through quantum error detection, which accepts an LBP only when its joint syndrome indicates no detected errors and discards it otherwise~\cite{glancy2006,pattison2025,maeda2025,marqversen2025}.

More advanced postselection uses the \textit{complementary gap} which quantifies the confidence of the decoder for error correction~\cite{bombin2024postselection,gidney2025yoked,smith2024mitigating,hetenyi2024entangled,sunami2025}.
We explain how the complementary gap is calculated. For each logical Bell observable $P\in\{XX,ZZ\}$, the same syndrome history can be explained by error patterns corresponding to either a $+1$ or $-1$ logical outcome. The decoder finds the minimum-weight error pattern for each outcome, with weights denoted by $W_P^{+}$ and $W_P^{-}$, respectively~\cite{bombin2024postselection,gidney2025yoked}.
The complementary gap for observable $P$ is the absolute difference between the two weights, which is defined in Equation~\eqref{eq:1}.
\begin{equation}
    g_P=\left|W_P^{+}-W_P^{-}\right|,
    \qquad g=\min(g_{XX},g_{ZZ}).
    \label{eq:1}
\end{equation}
This quantifies how strongly the decoder prefers one value over the other. A small gap indicates that the decoder may have difficulty distinguishing between the two logical outcomes because their decoding weights are similar. Leveraging this, postselection retains LBPs with gaps above a threshold ($g\ge g_{\mathrm{th}}$) and discards the rest.

\begin{figure}[htpb]
    \centering
    \vspace{-5pt}
    \includegraphics[width=\columnwidth]
        {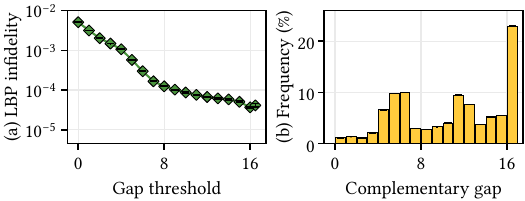}
        \vspace{-18pt}
       \caption{Distance-5 rotated surface-code LBP preparation with EB
    (Figure~\ref{fig:1}(c)) and five QEC rounds, using the same
    settings as in Section~\ref{sec:3}.
    (a) LBP infidelity after complementary-gap postselection and
    (b) gap distribution.}
    \label{fig:3}
    \vspace{-5pt}
\end{figure}

 Figure~\ref{fig:3} shows how complementary-gap-based postselection reduces the LBP infidelity after QEC rounds. At $g_{\mathrm{th}}=0$, all LBPs are retained, but the LBP infidelity is $5.1\times10^{-3}$.
 At $g_{\mathrm{th}}=16$, LBP infidelity falls to $3.7\times10^{-5}$. The gap distribution in Figure~\ref{fig:3}(b) shows that approximately $23\%$ of LBPs satisfy this cutoff, meaning that roughly one in four is accepted. Thus, complementary-gap postselection trades off LBP infidelity against the discarded LBPs.

\subsection{Requirements of LBP and Preparation Goals}
For reliable remote operations, an LBP must satisfy the following:

\begin{enumerate}[leftmargin=1.1em,labelsep=0.3em,nosep]
    \item An LBP must have a code distance at least comparable to that of the code block used for memory, because LER of remote operations is dictated by the block with the smallest code distance.
    \item LBP infidelity must be sufficiently low relative to that of the local operations, since the LER is accumulated  from all sources of error, including LBPs and local operations.
\end{enumerate}

\noindent Using a large-distance code increases spacetime volume per LBP generation. Consider an LBP encoded with the distance-$d$ rotated surface code. For each LBP half, $d^2$ data qubits and $d^2-1$ parity checks are required. So, $2(2d^2-1)$ physical qubits are needed in total. Once the LBP is generated, parity checks are repeatedly measured for multiple rounds to help distinguish data-qubit errors from measurement errors~\cite{dennis2002memory,fowler2012,horsman2012,tomita2014surface,stephens2014thresholds}. The \textit{spacetime volume} sums the physical qubits used in each QEC round.

As described in Section~\ref{sec:2.2}, a stricter postselection criterion (a higher $g_{\mathrm{th}}$) can reduce LBP infidelity but results in more discarded LBPs. Consequently, more PBPs and greater spacetime volume are required per delivered LBP.

\vspace{0.05in}
\noindent\textbf{Goal of LBP Preparation.} The goal of LBP preparation is to minimize PBP consumption and spacetime volume per delivered LBP while meeting the required code distance and target infidelity.

\subsection{Projection-Based LBP Preparation}

\begin{figure}[htpb]
    \vspace{-5pt}
\includegraphics[width=\columnwidth]{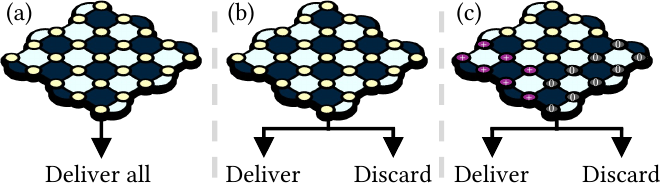}
\vspace{-15pt}
\caption{LBP preparation: (a) direct projection without postselection, (b) projection with postselection, and (c) EB with complementary-gap postselection. Each code block indicates an LBP half. Yellow dots denote PBP halves, while gray and  purple dots denote locally prepared states $|0\rangle$ and $|+\rangle$. Arrows indicate output selection after $d$ QEC rounds and decoding.}
    \label{fig:4}
    \vspace{-5pt}
\end{figure}

Prior work on projection-based LBP preparation uses fixed $d$ rounds of SMs and joint-syndrome error correction to convert PBPs into encoded LBPs~\cite{shi2025stabilizer}. LBPs can be retained without postselection~\cite{bonillaataides2025}, or selected using the number of estimated errors or complementary gap~\cite{maeda2025,sunami2025}. Figures~\ref{fig:4}(a) and (b) illustrate these two cases. We discuss other LBP distillation methods in Section~\ref{sec:8}.

\vspace{0.02in}
\noindent\textbf{Entanglement Boosting.} EB is a state-of-the-art projection-based LBP preparation method that reduces the PBP overheads via initialization~\cite{sunami2025}. As shown in Figure~\ref{fig:4}(c), PBP halves occupy only a smaller region of each LBP half, while the remaining data qubits are initialized in $|0\rangle$ or $|+\rangle$. These qubits together form the full target-distance blocks before the first QEC round. For target distance $d$, EB also performs $d$ QEC rounds on these blocks, computes the complementary gap, and retains LBPs with $g\ge g_{\mathrm{th}}$. Thus, EB reduces PBP overheads per delivered LBP while using complementary-gap postselection to exactly meet the target LBP infidelity.

\section{Limitations of the State-of-the-Art}\label{sec:3}

We explain the limitations of EB.
The data are collected via circuit-level simulation using an MWPM decoder~\cite{dennis2002memory,higgott2025sparseblossom}, SI1000 noise~\cite{gidney2021stim} with physical rate of $p=10^{-3}$ and a PBP infidelity of $p_{\mathrm{Bell}}=10^{-2}$.

\subsection{Fixed Rounds Limit Adaptive Stopping}
EB postselects LBPs after a fixed number of QEC rounds.

\begin{figure}[htpb]
    \centering
    \includegraphics[width=\columnwidth]
        {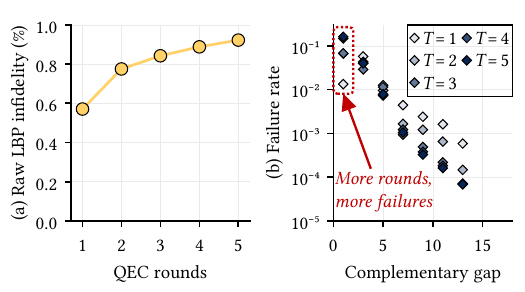}
    \vspace{-20pt}
    \caption{
    Fixed-round preparation with $3\times3$ corner PBPs
    and a $d$$=$$5$ surface code:
    (a) raw LBP infidelity and
    (b) failure rate.
    }
    \vspace{-7pt}
    \label{fig:5}
\end{figure}

 \vspace{0.05in}
 \noindent\textbf{Observation 1. More rounds can increase raw LBP infidelity.} Repeated QEC rounds help distinguish data-qubit errors from measurement errors~\cite{dennis2002memory}. Each additional round, however, also introduces more circuit noise. Figure~\ref{fig:5}(a) shows the LBP infidelity before postselection for different numbers of QEC rounds, with higher infidelity after more rounds. To explain this increase, Figure~\ref{fig:5}(b) groups failures by their final complementary gap.
 In the lowest-gap interval ($0\le g<2$), the failure rate rises from $1.33\%$ to $16.16\%$, which contributes an additional $0.45$ percentage points to raw infidelity.
 All other intervals ($g\ge2$) contribute $0.10$ percentage points less.
 Thus, \textit{the increase in low-gap failures outweighs the improvement in all other intervals combined}.

\vspace{0.05in}
\noindent\textbf{Key Insight 1. Discard low-gap LBPs early when further measurements offer little benefit.} Compute the complementary gap after each QEC round and assess the value of additional measurements. LBPs unlikely to benefit should be discarded immediately, avoiding the cost of completing the remaining fixed rounds.

\subsection{Large Codes Raise Discard Costs}

EB starts LBP preparation at the target code distance $d$.

\begin{figure}[b]
    \vspace{-5pt}
    \centering\includegraphics[width=\columnwidth]
        {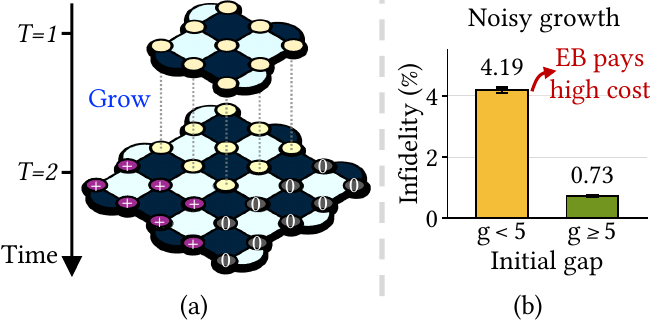}
    \vspace{-20pt}
        \caption{ Distance-3-to-5 growth with nine PBPs at a PBP error rate 0.01: (a) code expansion and (b) post-growth LBP infidelity and occurrence probability by first-round gap. }
    \label{fig:6}
\end{figure}

\vspace{0.05in}
\noindent\textbf{Observation 2. EB discards LBPs at a large code distance.}
Consider distance-5 LBP preparation with nine PBPs in EB as shown in Figure~\ref{fig:4}(c). Each QEC round costs 98 qubit-rounds across both QPUs, including for LBPs that are eventually discarded. To reduce the cost caused by a discard decision, we consider obtaining the first syndrome at a smaller code distance. The same nine PBPs can initialize a distance-3 LBP, reducing the spacetime volume per QEC round from 98 qubit-rounds to 34 qubit-rounds as shown in Figure~\ref{fig:6}(a). The LBP must still grow to distance 5 before delivery.

To use these first few QEC rounds for early discard, we need to know whether their complementary gaps indicate the risk of failure after growth. Growth introduces circuit noise, so we compare the gap after the first distance-3 round with the final LBP infidelity. For this comparison, all LBPs undergo the same noisy growth and subsequent QEC rounds.
After the first distance-3 round, $21.4\%$ of LBPs have $g<5$. As shown in Figure~\ref{fig:6}(b), these LBPs have a final infidelity of $4.19\%$, compared with $0.73\%$ for the remaining LBPs. Thus, the first distance-3 round already distinguishes LBPs with higher failure risk after growth. This provides an opportunity to discard such LBPs before incurring code growth.

\vspace{0.05in}
\noindent\textbf{Key Insight 2. Start at a smaller code distance and use the complementary gap to make discard decisions before incurring growth and target-distance QEC costs.}

\subsection{Design Requirements and Challenges}
Our insights impose two requirements: (1) use intermediate complementary gaps to decide whether to perform another QEC round, and (2) allow discard at a smaller code distance before the growth.

However, three challenges must be addressed: The current gap does not have information about whether each remeasurement or growth will reduce the LBP infidelity. We must also choose the initial code configuration, since a lower initial cost can be offset by additional QEC rounds or more discarded LBPs. Lastly, each decision must be quick to avoid extending the QEC cycle.

\section{Our Proposal: \design{}}
We propose \design{}, an LBP preparation compiler pass that answers the question \textit{``Should I Stay or Should I Grow?''} after each QEC round.

\subsection{Design Overview}

\begin{figure}[htpb]
    \vspace{-5pt}
\includegraphics[width=\columnwidth]{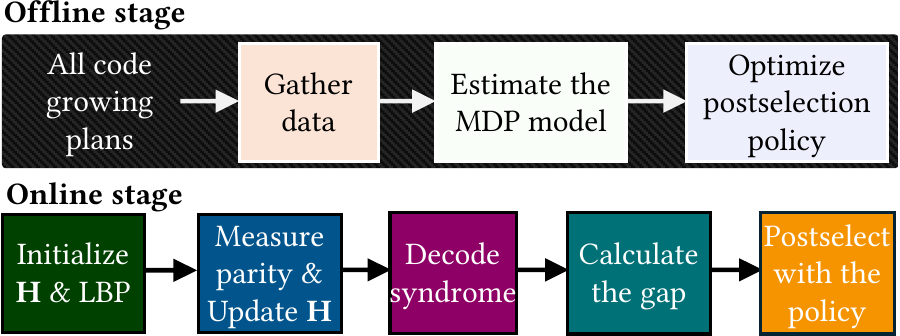}
    \vspace{-20pt}
\caption{Offline LBP preparation policy optimization and online policy deployment in \design{}.  (\textbf{H}: Syndrome history)}
\label{fig:7}
\end{figure}

As shown in Figure~\ref{fig:7}, \design{} is comprised of two stages: (1) offline policy optimization and (2) online policy deployment.
During the offline stage, \design{} collects circuit-level simulation data for each code growth plan. These data are used to build a Markov decision process (MDP) model that estimates how remeasurement and growth given the complementary gap affect subsequent postselection and LBP infidelity. \design{} then uses linear programming to jointly optimize the starting code and subsequent postselection decisions, balancing PBP overheads and spacetime volume under the target infidelity constraint and target code distance.

During the online stage, \design{} is deployed on a runtime controller. For each LBP preparation attempt, the controller selects a growth plan using the optimized policy and initializes the LBP at the starting code distance of that plan. The controller records the outcomes of the first QEC round as the initial syndrome history and appends the outcomes of subsequent rounds. After each QEC round, the MWPM decoder decodes the accumulated syndrome history and computes the complementary gap. The controller then applies the optimized policy to decide whether to remeasure at the current code distance, grow the code before the next QEC round, discard the attempt, or deliver the LBP at the required distance.

    \subsection{MDP Formulation for LBP Preparation}

\begin{figure}[htpb]
    \centering\includegraphics[width=\columnwidth]{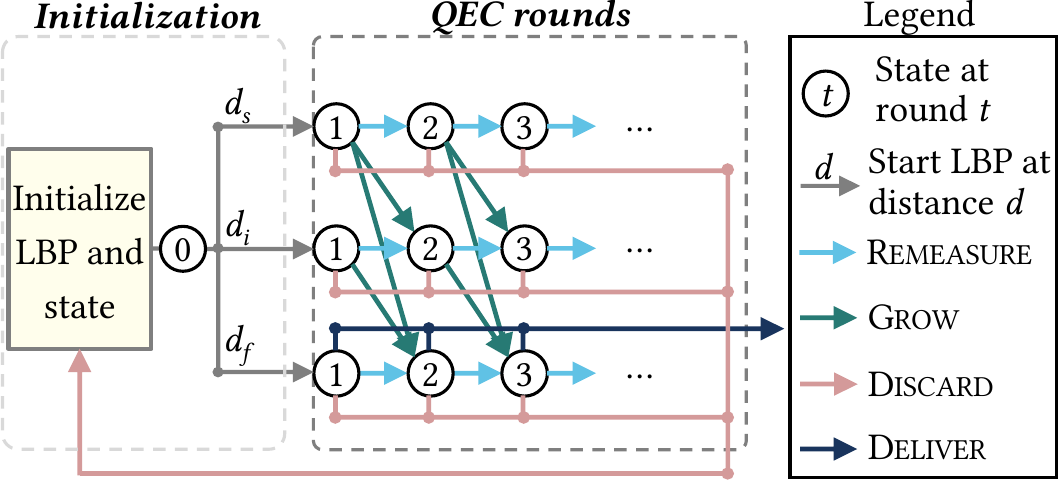}
    \vspace{-20pt}
    \caption{MDP overview for adaptive LBP preparation. Nodes group preparation states by code distance and the total number of completed QEC rounds. Edges indicate initial code selection and subsequent preparation actions. The code growth plan, round counts at each distance, and current and previous complementary-gap intervals are omitted for clarity.}
    \label{fig:8}
    \vspace{-15pt}
\end{figure}

 We formulate LBP preparation as a constrained MDP, as illustrated in Figure~\ref{fig:8}. After initial code selection, the policy selects an action based on the preparation state after each QEC round. The \textsc{Discard} decision restarts LBP preparation, while \textsc{Deliver} returns the LBP at the target distance.

\noindent\textbf{LBP Preparation State.}
A code growth plan specifies the initial code configuration and the code growth allowed during preparation. Let $c[t]$ identify the selected code growth plan and the code configuration after round $t$, including the code distance, PBP locations, local initial states, and the code growth completed by round $t$. The vector $r[t]=(r_{d_s}[t],\ldots,r_{d_f}[t])$ records the number of completed QEC rounds at each distance, where $d_s$ and $d_f$ are the initial and final distances. For a plan that grows from distance 3 to 5 to 7, the total number of completed rounds is $t=r_3[t]+r_5[t]+r_7[t]$.

Let $s[t]$ denote the syndrome obtained in round $t$, and let $s[1:t]$ denote the accumulated syndrome history. Decoding this history gives the current complementary gap $g[t]$. We compute the gap as the difference between the smallest and second-smallest MWPM weights among the four logical Bell classes: no error, $X$ error, $Z$ error, and $Y$ error. This value equals $g$ in Equation~\eqref{eq:1}. The decoder computes the current complementary gap from the accumulated syndrome history. The previous complementary gap $g[t-1]$ is computed from $s[1:t-1]$.

We group complementary-gap values into intervals to limit the number of MDP states. As a rough illustration, for fixed configuration, round counts, and action, using 0.1-wide intervals for the previous, current, and next gaps over $[0,30)$ expands the possible transition entries from $4\times30^2$ under our discretization to $300^3$, a $7{,}500\times$ increase. If this expansion applied throughout the model, maintaining the average observations per entry would scale a $2.3\times10^8$-shot training budget to \underbar{\textit{approximately $1.7\times10^{12}$ shots}}. For the previous complementary gap, we use four intervals: $[0,6)$, $[6,10)$, $[10,14)$, and $[14,\infty)$. The values $B(g[t])$ and $B_{\mathrm{prev}}(g[t-1])$ identify the intervals containing the current and previous complementary gaps, respectively. At $t=1$, we use a separate value to indicate that no previous gap exists. The state after QEC round $t$ is

\begingroup
\vspace{-5pt}
\addtolength{\abovedisplayskip}{-5pt}
\addtolength{\abovedisplayshortskip}{-5pt}
\addtolength{\belowdisplayskip}{-5pt}
\addtolength{\belowdisplayshortskip}{-5pt}
\begin{equation}
 \textstyle
 x[t]:=\big(c[t],\,r[t],\,B(g[t]),\,
 B_{\mathrm{prev}}(g[t-1])\big).
 \label{eq:2}
\end{equation}
\endgroup

\vspace{0.05in}
\noindent\textbf{Set of Postselection Decisions.}
The target distance and maximum number of rounds are specified for each set of preparation requirements. After each round, \design{} postselects as follows:

\begin{itemize}[leftmargin=1.1em,labelsep=0.3em,nosep]
    \item \textsc{\textbf{Remeasure}}: Run a QEC round at the current code distance.
    \item \textsc{\textbf{Grow}}: Expand both blocks under the selected growth plan \textit{without adding PBPs}, then run the next round at the new distance.
    \item \textsc{\textbf{Discard}}: End the current attempt without delivery and start a new attempt with a newly selected plan.
    \item \textsc{\textbf{Deliver}}: Return the LBP and complete the request. This action is permitted only at the required code distance.
\end{itemize}
\noindent We define $\mathcal{A}$ as the set of \textsc{Remeasure} and \textsc{Grow} actions. Both require at least one remaining QEC round.

 \noindent\textbf{Empirical Transition Model.} We use circuit-level simulation data to precisely estimate how remeasurement and growth change the preparation state. The training simulations require hundreds of millions of shots. Given the target distance and PBP error rate $p_{\mathrm{Bell}}$, we collect $2.3\times10^8$ shots of data once and reuse them to estimate the model. For code growth plan $i$, $\mu_i(x)$ gives the probability of reaching state $x$ after the first QEC round. For $a\in\mathcal{A}$, let $n_a(x,x')$ count the transitions from $x$ to $x'$ observed after action $a$ in the collected data. We estimate the transition probability from the fraction of transitions from $x$ under action $a$ that reach $x'$:

\vspace{-4pt}
\begingroup
\addtolength{\abovedisplayskip}{-3pt}
\addtolength{\abovedisplayshortskip}{-3pt}
\addtolength{\belowdisplayskip}{-3pt}
\addtolength{\belowdisplayshortskip}{-3pt}
\begin{equation}
 \widehat P_a(x'\mid x)
 =n_a(x,x')\big/\textstyle{\sum_u n_a(x,u)}.
\end{equation}
\endgroup

If the denominator is zero, the model assigns the continuation
to a separate unavailable-gap state, where delivery is prohibited.

To estimate failure risk, let $n(x)$ count training attempts reaching $x$, and let $f(x)$ count those with an incorrect logical $XX$ or $ZZ$ correlation after a separate offline error-correction step that introduces no new faults. Equation~\eqref{eq:4} gives the Jeffreys posterior mean of the logical-failure probability conditioned on $x$.

\vspace{-4pt}
\begingroup
\addtolength{\abovedisplayskip}{-3pt}
\addtolength{\abovedisplayshortskip}{-3pt}
\addtolength{\belowdisplayskip}{-3pt}
\addtolength{\belowdisplayshortskip}{-3pt}
\begin{equation}
 \widehat p_{\mathrm{fail}}(x)
 =(f(x)+0.5)\big/(n(x)+1).
 \label{eq:4}
\end{equation}
\endgroup

\begin{figure}[htpb]
    \vspace{-8pt}
    \centering
    \includegraphics[width=\linewidth]{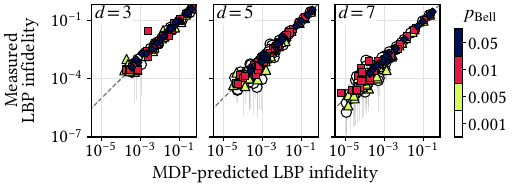}
    \vspace{-23pt}
    \caption{Strong correlation between predicted and measured LBP infidelities with 12 configurations with $\rho$$\in$$[0.906,0.994]$.}
    \label{fig:9}
    \vspace{-15pt}
\end{figure}

Figure~\ref{fig:9} compares the predicted LBP infidelity for each preparation state with
estimates from independent circuit-level simulations of tuples $(p_{\mathrm{Bell}},~d)\in\{0.001,0.005,0.01,0.05\}\times\{3,5,7\}$. Across twelve fixed-round conditions, the predictions
and measurements are strongly correlated (Spearman's coefficient $\rho\in[0.906,0.994]$) among states with logical failures observed in these simulations.

\subsection{Postselection Policy Optimization}

 \vspace{0.05in} \noindent\textbf{Objective.} The policy minimizes the modeled cost per delivered LBP subject to a constraint on predicted infidelity. Costs include resources consumed by preparations that are discarded. We combine PBP consumption $M$ and spacetime volume $V$ as $M+wV$, where $w$ sets their tradeoff coefficient. Let $C_0(i)$ denote the cost of reaching the first decision state of plan $i$, and let $C(x,a)$ denote the additional cost of taking action $a$ at state $x$.

 \vspace{0.05in}
\noindent\textbf{Policy Optimization.} We optimize the expected number of plan starts and action selections needed for one delivered LBP. Let $z_i$ be the expected number of times code growth plan $i$ is selected, and let $y_{x,a}$ be the expected number of times action $a$ is taken at state $x$, both per delivered LBP. The optimization problem~\eqref{eq:5} minimizes the expected cost by multiplying these counts by $C_0(i)$ and $C(x,a)$, respectively, subject to the target infidelity $\varepsilon$. We set $y_{x,a}=0$ for actions that are not allowed at state $x$.

\newpage
~\\[-18pt]
\begin{equation}
    \begin{aligned}
    \textstyle{\min_{y,z\ge0}~}&
     \textstyle{J(y,z) := \sum_i z_iC_0(i)+\sum_{x,a}y_{x,a}C(x,a)}
     \\[-2pt]
    \textstyle{\mathrm{s.t.}}\quad&
     \textstyle{\sum_a y_{x,a}
     -\sum_{x'}\sum_{a'\in\mathcal{A}}
     y_{x',a'}\widehat P_{a'}(x|x')
     =\sum_i z_i\mu_i(x),}
     \\[-2pt]
    &\textstyle{\sum_x y_{x,\mathrm{deliver}}=1,\quad
     \sum_x y_{x,\mathrm{deliver}}
     \widehat p_{\mathrm{fail}}(x)\le\varepsilon,}
    \end{aligned}
    \label{eq:5}
\end{equation}
\vspace{-5pt}

The constraints preserve attempt flow, normalize one expected
delivery, and bound its predicted failure probability.
The solution selects each new plan with probability
$z_i/\sum_j z_j$.
At states with positive flow, the action probabilities are
$\pi(a\mid x)=y_{x,a}/\sum_b y_{x,b}$.
During preparation, each new syndrome-extraction round updates
$x[t]$, and the policy selects the next action from
$\pi(\cdot\mid x[t])$.

\begin{figure}[htpb]
    \vspace{-5pt}
    \centering    \includegraphics[width=\columnwidth]{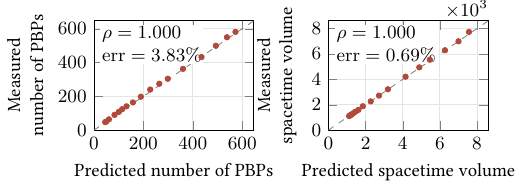}
    \vspace{-23pt}
    \caption{Predicted and independently measured PBP and spacetime costs per delivered LBP for 17 fixed policies per objective. Costs include discarded attempts.}
    \vspace{-10pt}
    \label{fig:10}
\end{figure}

Figure~\ref{fig:10} compares predicted PBP consumption and spacetime volume per delivered LBP with measurements from independent circuit-level simulations. For each objective, we evaluate 17 policies fixed before these simulations. Predicted and measured costs have identical rankings for both objectives, with Spearman correlation $\rho=1$. Mean absolute relative errors are only $3.83\%$ for PBP consumption and $0.69\%$ for spacetime volume.

\subsection{Deployment of Policy to Controller}
\begin{figure}[b]
    \centering
    \vspace{-5pt}
    \includegraphics[width=\columnwidth]{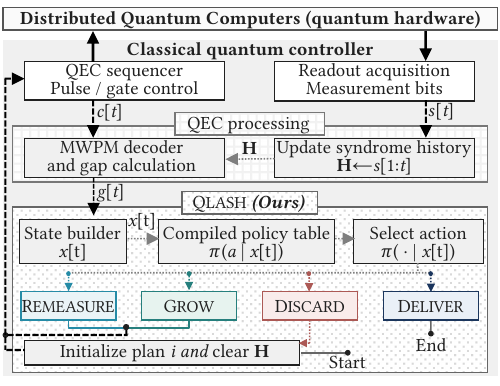}\vspace{-15pt}
    \caption{Online control of \design{}.}
    \label{fig:11}
\end{figure}
Figure~\ref{fig:11} shows how \design{} connects to the QEC sequencer within the classical controller. After each QEC round, readout acquisition collects the measurement results and passes $s[t]$ to the history-update block, which appends it to the syndrome history $\textbf{H}$. The controller increments the round count for the code distance used in that round. The MWPM decoder computes the complementary gap $g[t]$ using the accumulated history $\textbf{H}=s[1:t]$ and the code configuration $c[t]$ supplied by the QEC sequencer. The state builder forms $x[t]$ from $c[t]$, the updated round counts, and the current and previous gap intervals. For the first round, a separate value indicates that no previous gap exists. \design{} reads the policy entry for $x[t]$. The action-selection block selects an action according to $\pi(\cdot\mid x[t])$, sampling when multiple actions have nonzero probability. For \textsc{Remeasure}, the QEC sequencer runs another QEC round at the current code distance. For \textsc{Grow}, it expands both code blocks as permitted by the selected plan and then runs the next round at the new distance. Both actions retain $\textbf{H}$ and the completed round counts, so decoding after growth includes the syndrome measurements obtained before growth. The next decision uses $B_{\mathrm{prev}}(g[t])$ as the previous-gap component. \textsc{Discard} repeats plan selection and initialization to start a new LBP preparation. \textsc{Deliver} returns the LBP at the required code distance and completes preparation.

\begin{figure}[b]
    \centering
    \vspace{-30pt}
    \includegraphics[width=\columnwidth]{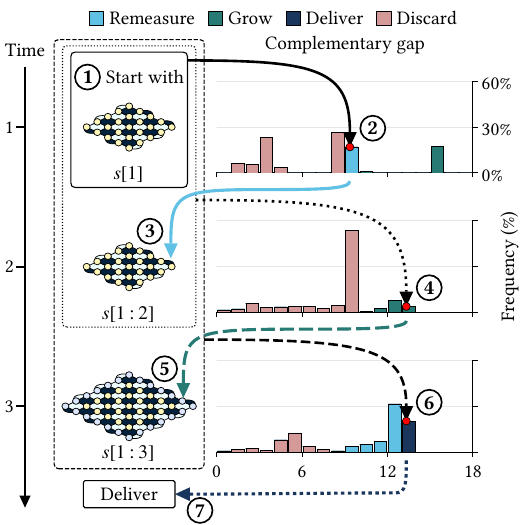}
    \vspace{-22pt}
    \caption{Example of online preparation of a distance-7 LBP with \design{}. A red dot indicates the complementary gap value of each round. Bar heights show complementary-gap frequencies at each round, and colors denote actions. \design{} remeasures at distance~5, grows both blocks to distance~7, and delivers the pair after decoding all three rounds.}
    \label{fig:12}
\end{figure}

\vspace{0.05in}
\noindent\textbf{Postselection via \design{}.} Figure~\ref{fig:12} shows one preparation of a distance-7 LBP.
\circledtext{1}~\design{} samples a code growth plan that starts at distance~5. Each QPU performs the first QEC round for its LBP half.
\circledtext{2}~The decoder computes the complementary gap from $s[1]$, and \design{} selects \textsc{Remeasure} from the gap distribution for $x[1]$. \circledtext{3}~Each QPU again performs a second QEC round on the LBP. \circledtext{4}~The decoder updates the gap using $s[1:2]$, and \design{} selects \textsc{Grow} for $x[2]$.
\circledtext{5}~Each QPU expands its block to distance~7 and performs the third QEC round. \circledtext{6}~The decoder computes the new gap from all three rounds, including the two before growth. The gap falls in the same interval as in round two, but delivery is now permitted because the blocks have reached distance~7. \design{} samples \textsc{Deliver} from the entry for $x[3]$. \circledtext{7}~\design{} returns the LBP.

\clearpage
\section{Evaluation Methodology}

\subsection{Research Questions}

We evaluate whether \design{} reduces the resources required
to deliver an LBP under a common code-distance requirement
and infidelity limit. We address four research questions via evaluation:

\vspace{0.05in}
\begin{itemize}[leftmargin=1.1em,labelsep=0.3em,nosep]

    \item \textbf{\textit{RQ1: Resource efficiency?}}  We compare PBP consumption and active spacetime per delivered LBP at the same destination code distance and matched measured delivered-LBP infidelity.\\[-9pt]

    \item \textbf{\textit{RQ2: Sources of benefit?}} We quantify how much syndrome-adaptive stopping
    and conditional code growth each contribute to resource savings.\\[-9pt]

    \item \textbf{\textit{RQ3: Operating regimes?}} We vary PBP infidelity, code distance, and target LBP infidelity
    to identify when \design{} retains its resource advantage over the baselines.\\[-9pt]

    \item \textbf{\textit{RQ4: Classical overheads?}} We measure the offline compilation cost and online overhead of \design{}.
    We also report how much \design{} saves after accounting for decision latency.
\end{itemize}

\subsection{Figures of Merit}

We define LBP infidelity as the probability that a delivered pair has an incorrect logical $XX$ or $ZZ$ correlation. To compare methods at matched output quality, we hold the destination code distance fixed and match measured delivered-LBP infidelity across methods in each comparison. We report the following resource costs per delivered LBP.

\vspace{0.05in}
\begin{itemize}[leftmargin=1.1em,labelsep=0.3em,nosep]
    \item\textbf{PBP Consumption.}
    We count PBPs consumed by every attempt, including discarded attempts, per delivered LBP. \\[-9pt]

    \item\textbf{Spacetime Volume.} We count the occupied physical-qubit time of data qubits and syndrome ancillas across both QPUs, including time spent on discarded attempts, per delivered LBP.
\end{itemize}

\vspace{0.05in}
\textit{Ideally, low PBP consumption and spacetime volume are better.}

\subsection{Codes and Noise Model}

\vspace{0.05in}
\noindent\textbf{Code.} We evaluate \design{} with the rotated surface code. We use the CNOT order of the rotated-surface-code circuits generated by Stim~\cite{gidney2021stim}, which keeps hook errors from reducing the code distance.

  \vspace{0.05in}  \noindent\textbf{PBP Noise and Initialization.} We set the input PBP infidelity to $p_{\mathrm{Bell}}=0.05$ for the main comparison, which is more than 10$\times$ the physical error rate. To isolate the effect of photonic interconnect noise, we vary $p_{\mathrm{Bell}}$ from $10^{-3}$ to $10^{-1}$ while keeping $p$ fixed. Each PBP starts in $\lvert\Phi^+\rangle$. Before the first local operation, we apply $X$, $Y$, or $Z$ to one half with probability $p_{\mathrm{Bell}}/3$ each~\cite{sunami2025,bonillaataides2025}.

  We initialize the remaining data qubits with a noisy $Z$-basis reset and apply a noisy Hadamard wherever the preparation plan requires $\lvert+\rangle$. We omit idle noise on each data qubit before its first local operation. Syndrome-ancilla resets are noisy throughout preparation.

 \vspace{0.05in} \noindent\textbf{Circuit-Level Noise.} We conduct circuit-level simulations with faults in local gates, syndrome-ancilla resets, measurements, and idle steps during syndrome extraction and code growth. We set the base physical error rate to $p=10^{-3}$ and apply the SI1000 component error probabilities~\cite{gidney2021honeycombmemory,gidney2022honeycomb,gidney2024inplace,gidney2025yoked}. We summarize the noise of each component in Table~\ref{tab:1}.

\begin{table}[htpb]
\centering
\caption{SI1000 circuit noise model used in this paper.}
\vspace{-10pt}
\label{tab:1}
\setlength{\tabcolsep}{3pt}
\begin{tabularx}{220pt}{|l|X|}
\hline
\textbf{Operation} & \textbf{Error channel} \\
\hline
CNOT & Two-qubit depolarization, $p$ \\
Single-qubit Clifford & Depolarization, $p/10$ \\
$Z$-basis reset & $X$ error after reset, $2p$ \\
$Z$-basis measurement & $X$ error before readout, $5p$ \\
Idle in each step & Depolarization, $p/10$ \\
Idle during reset/readout & Additional depolarization, $2p$ \\
\hline
\end{tabularx}
\vspace{-10pt}
\end{table}

\vspace{0.05in}
\noindent\textbf{Decoding and Evaluation.}
At round $t$, we compute the EB complementary gap $g[t]$ from the syndrome history $s[1:t]$ using Sunami et al.~\cite{sunami2025}'s four bases ($|00\rangle,\,|01\rangle,\,|10\rangle,\,|11\rangle$) of the MWPM matching procedure~\cite{sunami2025,ebCode}. We assume that the latency to calculate the complementary gap is negligible. We conduct a separate sensitivity study of the complementary-gap latency in Section~\ref{sec:6.6}.

\subsection{Hardware Assumption}
 \noindent\textbf{PBP Supply.} We assume that the link supplies undistilled PBPs, with one qubit of each pair at each QPU~\cite{stephenson2020,saha2025,main2025,magnard2020}. The parameter $p_{\mathrm{Bell}}$ denotes the infidelity of each pair when it enters LBP preparation. We count each input pair once and exclude the time and resources used to generate it. Code growth uses only local operations and consumes no additional PBPs~\cite{horsman2012,marqversen2025}.

\vspace{0.05in}
\noindent\textbf{Space for Code Growth.} Starting LBP preparation at a smaller code distance leaves some qubits needed at the target distance unused during the early QEC rounds. Reserving these qubits from the start would prevent other operations from using them. We assume that other operations can use them beforehand and release them without delaying growth. The growth circuit then resets the released qubits before adding them to the LBP patches. Qubit reset and reuse have been demonstrated experimentally~\cite{magnard2018reset,finkelstein2024reuse}. Under this assumption, we report active spacetime, which counts the physical-qubit time occupied by LBP preparation.
\subsection{Baselines}
We compare \design{} against the following methods at the same destination distance and apply the same noise model.

\vspace{0.05in}
\noindent\textbf{Entanglement Boosting.}
Our EB baseline follows the published protocol, with the authors' public implementation as the reference for preparation and complementary-gap decoding~\cite{sunami2025,ebCode}.
It places PBPs in a corner, prepares the remaining data qubits locally, performs $d$ QEC rounds at the destination distance, and applies a final gap threshold.
Code projection encodes PBPs at all $d^2$ data qubits.

\vspace{0.05in}
\noindent\textbf{\design{} Variants.}
To isolate each optimization  (per-round delivery decision and code growth), we compare \design{} against its ablation that adds syndrome-adaptive remeasurement, discarding, and delivery decisions with the code at the target distance.

\subsection{Experimental Setup}
 We run Python Monte Carlo simulations on compute clusters using Stim~\cite{gidney2021stim} and PyMatching~\cite{higgott2025sparseblossom} for circuit simulation and MWPM decoding. We evaluate the controller through post-route analysis and simulation in Vivado 2025.2~\cite{amd2025vivado}, targeting an FPGA at 200\,MHz. We will release the source code, all experiment configurations with random seeds, and analysis scripts.
\clearpage

\section{Evaluation}

\subsection{How \design{} Reduces LBP Preparation Cost}

\noindent \textbf{Setup.} We compare three policies for preparing a distance-7 LBP at a local error rate of $p=10^{-3}$. EB measures every attempt for seven rounds on the destination patch, then uses decoder confidence to decide whether to deliver the pair. \design{}-Opt1 uses the same patch but can stop after any round. \design{}-Opt2 can also start on a smaller patch and use the syndrome to decide whether to grow. We count PBPs and active qubit-time over all attempts, including discarded ones, per delivered LBP. At each $p_{\mathrm{Bell}} \in \{0.025, 0.05, 0.06\}$, we match delivered-LBP infidelity: point estimates for EB and \design{} lie in $[1.97, 2.00]\times10^{-5}$ and are statistically indistinguishable, each based on $10^{4}$ logical failures (95\% confidence intervals within $\pm2\%$).

\begin{figure}[htpb]
    \centering
\vspace{-5pt}    \includegraphics[width=\columnwidth]{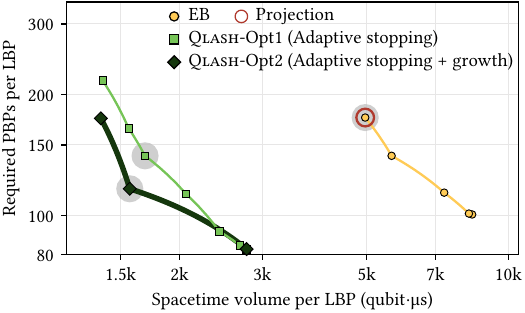}
\vspace{-15pt}
    \caption{PBP consumption and active spacetime per delivered distance-7 LBP at matched delivered-LBP infidelity. The policy-selection target is $2\times 10^{-5}$. Gray-shaded points mark costs at tradeoff coefficient $w = 0.122$.}
    \label{fig:13}
\vspace{-5pt}
\end{figure}

   \vspace{0.05in}
   \noindent \textbf{Resource Efficiency.}     \design{} uses \textbf{\underbar{18.8--25.2\% fewer PBPs}} and \textbf{\underbar{19.6--68.9\% less active spacetime}} per delivered LBP than EB. Figure~\ref{fig:13} illustrates these savings at additional operating points with matched delivered-LBP infidelity. \design{}-Opt1 uses the same distance-7 block as EB yet consumes about 62\% less active spacetime per delivery. The two policies begin to differ after the first QEC round.

\begin{figure}[htpb]
    \centering
    \includegraphics[width=\columnwidth]{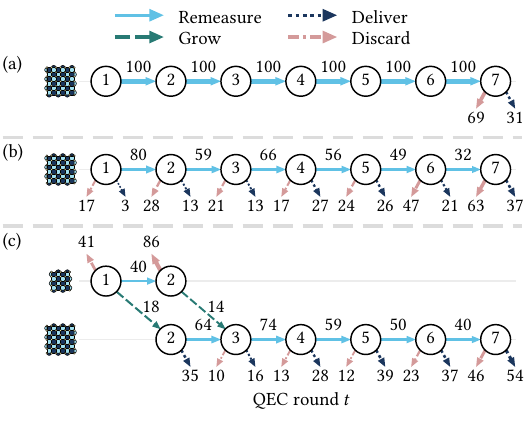}
    \vspace{-20pt}
    \caption{Preparation decisions by round for (a) EB with 7 QEC rounds, (b) \design{}-Opt1, and (c) \design{}-Opt2. Each edge indicates the percentage of attempts at its source node that take the indicated action.}
    \label{fig:14}
\end{figure}

\noindent\textbf{Takeaway~1. Choose which attempts receive another round.}
Figure~\ref{fig:14} traces one component of each highlighted policy. In Figure~\ref{fig:14}(a), EB measures every attempt for seven rounds, then discards 68.7\% of them. In Figure~\ref{fig:14}(b), \design{}-Opt1 can already deliver or discard after round one: 3\% are delivered, 17\% are discarded, and 80\% continue. Of those that reach round two, 59\% are measured again. Thus, about 53\% end within two rounds, while some continue through round seven.

\vspace{0.05in}
 \noindent\textbf{Takeaway~2. Decide which attempts need the larger block.} Even when \design{}-Opt1 discards an attempt after one round, it has occupied the full distance-7 block on both QPUs for that round. In Figure~\ref{fig:14}(c), \design{}-Opt2 makes its first decision on a smaller block. Of attempts in this 25-PBP component, 18\% grow after the first round and 40\% are measured again on the smaller block. Only 14\% of that second group then grow. In all, about one attempt in four reaches the destination block.

\noindent\textbf{Can EB with a Smaller Fixed Number of Rounds Recover the Savings?} If \design{}-Opt1 often stops within two rounds as shown in Figure~\ref{fig:14}(b), why not fix the QEC round count at one or two?

\begin{figure}[htpb]
    \centering

    \vspace{-5pt}
    \includegraphics[width=\columnwidth]{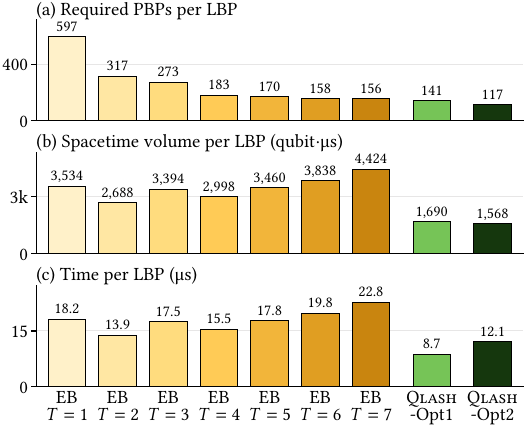}
    \vspace{-12pt}
    \caption{Fixed schedules of one through seven rounds compared
    with \design{}-Opt1 and \design{}-Opt2. (a) PBP
    consumption, (b) active spacetime, and (c) preparation time per
    delivered LBP, including discarded attempts. Delivered-LBP infidelity point estimates range from $1.97\times10^{-5}$ to $2.00\times10^{-5}$.}
    \label{fig:15}
\end{figure}

 To answer this question, Figure~\ref{fig:15} compares fixed schedules of one through seven rounds. For each round count, we tune the final delivery rule using the current gap and, when available, the previous gap. The fixed policies can therefore use the complementary gap to decide which pairs to deliver. However, they must choose the stopping round before observing any syndrome.

 Among the fixed policies shown, stopping after one round costs 597 PBPs per delivery, the highest PBP cost in Figure~\ref{fig:15}(a). Two rounds minimize active spacetime at 2,688 qubit-$\mu$s per delivery as shown in Figure~\ref{fig:15}(b), but consume 317 PBPs. Seven rounds minimize PBP consumption at 156.4 per delivery, but use 4,424 qubit-$\mu$s. \design{}-Opt1 uses 141 PBPs and 1,690 qubit-$\mu$s per delivery, below the best fixed result for each resource.

 \vspace{0.05in} \noindent\textbf{Why does PBP consumption per delivery fall?} Stopping early avoids later QEC rounds, but cannot recover the PBPs spent at the start of an attempt. The PBP cost in Figure~\ref{fig:15}(a) is the average number of PBPs per attempt divided by the fraction of attempts delivered. EB and \design{}-Opt1 both start with 49 PBPs. Across the full policies, \design{}-Opt1 delivers 34.8\% of attempts, compared with 31.3\% for EB. This higher delivery rate lowers PBP consumption from 156.4 to 141 per delivery. \design{}-Opt2 saves PBPs for a different reason. Its delivery rate falls from 34.8\% with \design{}-Opt1 to 19.6\%, but it starts with only 22.85 PBPs per attempt on average. The lower initial cost outweighs the additional retries, reducing PBP consumption to 117 per delivery, 17.2\% below \design{}-Opt1.

 \vspace{0.05in} \noindent\textbf{Why does \design{}-Opt2 save spacetime despite taking longer?} Time per delivery rises from 8.7~$\mu$s with \design{}-Opt1 to 12.1~$\mu$s with \design{}-Opt2 as shown in Figure~\ref{fig:15}(c). Dividing active spacetime per delivery in Figure~\ref{fig:15}(b) by time per delivery in Figure~\ref{fig:15}(c) gives an average of $1{,}690/8.7$$\approx$$194$ active qubits for \design{}-Opt1, compared with $1{,}568/12.1$$\approx$$130$ for \design{}-Opt2. The smaller footprint offsets the extra time, cutting active spacetime by 7.2\%.

 \subsection{Is a Fixed Growth Round Sufficient?}
We test whether preselected growth schedules can recover the benefits of conditional growth. Fixed grows surviving attempts after a predetermined round. We report the best and worst policies over $T=1$--$3$ by weighted cost. Mixed samples a fixed-growth policy before each attempt, applying the infidelity criterion to the combined output. Both baselines retain syndrome-based early discard.

All methods start with distance-5 blocks and 25 PBPs and deliver distance-7 LBPs at matched infidelity. At $p_{\mathrm{Bell}} = 0.025$ and $w = 0.122$, delivered-LBP infidelity point estimates span $[3.58, 3.60]\times10^{-5}$. Mixed varies schedules across attempts but cannot revise assigned growth rounds. \design{} uses each syndrome to choose between remeasurement and growth. Growing early increases the footprint of later rounds, while delaying growth reduces it but can postpone delivery. Table~\ref{tab:2} shows that \design{} uses less active spacetime per attempt and delivers more often than Mixed. Since each attempt consumes 25 PBPs, higher yield reduces PBP consumption per delivery. Together with lower spacetime per attempt, it also reduces active spacetime per delivered LBP.

\begin{table}[b]
\vspace{-10pt}
    \centering
    \captionsetup{skip=5pt}
    \caption{Performance comparison against \design{} variants. A metric
    $J$ is the weighted cost per delivered LBP.
    Spacetime volume per attempt (ST/attempt) is measured in qubit$\cdot$rounds.}
    \label{tab:2}

\setlength{\tabcolsep}{2.5pt}
\begin{tabularx}{240pt}{|>{\raggedright\arraybackslash}X|r|r|r|}
\hline
\textbf{Method} & $J$ $\downarrow$
& \textbf{ST/attempt} $\downarrow$
& \textbf{Yield (\%)} $\uparrow$ \\
\hline
Worst Fixed ($T=1$) & 414.20 & 643.99 & 25.00 \\
Best Fixed ($T=3$)  & 290.91 & 535.30 & 31.04 \\
Mixed              & 244.34 & 401.58 & 30.28 \\\hline
\design{} (Adaptive Growth)         & \textbf{217.69}
                   & \textbf{357.38}
                   & \textbf{31.51} \\
\hline
\end{tabularx}
\end{table}

\subsection{Operating Regimes of Adaptive Preparation}

 PBP infidelity varies across quantum interconnects. For reference, experiments have reported Bell-pair infidelities of approximately $1.1\%$ between microwave-connected superconducting QPUs~\cite{niu2023interconnects} and $3\%$--$6\%$ between photonically connected trapped-ion QPUs~\cite{main2025,stephenson2020}. We therefore evaluate \design{} with $p_{\mathrm{Bell}}\in[0.001,0.1]$.

\begin{figure}[htpb]
     \centering
     \vspace{-10pt}
     \includegraphics[width=\columnwidth]{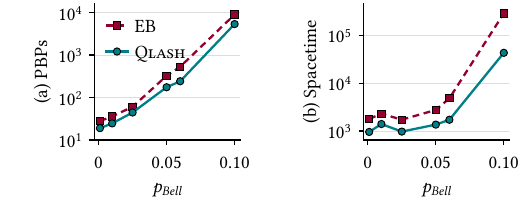}
     \vspace{-20pt}
     \caption{Impact of PBP infidelity ($p_{Bell}$) on (a) PBPs and (b) active spacetime ($d$$=$$7$, $p$$=$$10^{-3}$, target LBP infidelity $2\times10^{-5}$).}
     \label{fig:16}
     \vspace{-10pt}
 \end{figure}

Across all six operating points, \design{} consistently outperforms EB. Figure~\ref{fig:16} shows reductions of $26.5$--$54.6\%$ in PBP consumption and $39.3$--$85.0\%$ in active spacetime per delivered LBP, using the same policy pair for both metrics at each point. The largest spacetime reduction occurs at $p_{Bell}=0.10$, where \design{} reduces spacetime by $85\%$ while consuming $40\%$ fewer PBPs than EB.

\subsection{Compilation Time and Runtime Overheads}
\label{sec:6.4}
We evaluate the classical overheads of \design{}.

\vspace{0.05in}
\noindent\textbf{Offline Compilation.}
Given an estimated MDP model, the mean compilation time per LP call is 3.39\,s for $d_f=7$ and 16.88\,s for $d_f=9$, as shown in Table~\ref{tab:3}(a). These times cover model loading through policy serialization and exclude the circuit-level simulations used to collect training data. Peak batch memory is 259.2\,MiB and 513.6\,MiB, respectively. The resulting policy is reused across LBP preparations under the same conditions. If the cost weight or target infidelity changes, \design{} optimizes a new policy using the saved model.

\vspace{0.05in}
\noindent\textbf{Online Latency and Hardware Resources.}
After each QEC round, the FPGA controller reads the policy entry for $x[t]$ and selects an action according to $\pi(\cdot\mid x[t])$, sampling when multiple actions have nonzero probability (Figure~\ref{fig:11}). Simulation of the synthesized distance-7 controller shows that the selected action is output five clock cycles after input acceptance, as shown in Table~\ref{tab:3}(b). Timing analysis of the FPGA implementation supports operation at 200\,MHz, giving a \underbar{\textbf{decision latency of \textit{25\,ns}}}. The implementation uses 329 LUTs and 131 flip-flops, with no DSPs. These results show that the compiled \design{} policy supports per-round action selection with low latency and hardware overhead.
\begin{table}[b]
\vspace{-20pt}
\centering
\captionsetup{skip=5pt}
\caption{Offline compilation and FPGA policy evaluation.}
\label{tab:3}

\small
\setlength{\tabcolsep}{3pt}
\renewcommand{\arraystretch}{1.1}
\begin{tabularx}{\columnwidth}{@{}|X|r|r|@{}}
\hline
\textbf{(a) Offline compilation} & $d_f=7$ & $d_f=9$ \\
\hline
LP states / variables & 14,208 / 39,618 & 29,748 / 84,502 \\
Mean wall time per LP & 3.39\,s & 16.88\,s \\
Peak batch memory & 259.2\,MiB & 513.6\,MiB \\
\hline
\multicolumn{3}{|l|}{
    \textbf{(b) Distance-7 FPGA controller at 200\,MHz}
} \\
\hline
\multicolumn{3}{|c|}{
    \includegraphics[
        width=.97\columnwidth,
        trim=0 0 0.1pt 0,
        clip
    ]{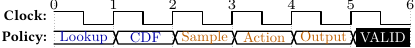}
} \\
\hline
\multicolumn{3}{|c|}{%
    \footnotesize
    \setlength{\tabcolsep}{2pt}%
    \begin{tabular*}{0.96\columnwidth}{
        @{\extracolsep{\fill}}c|c|c|c|c@{}
    }
    \textbf{Latency}: 25\,ns &
    \textbf{LUT}: 329 &
    \textbf{FF}: 131 &
    \textbf{BRAM}: 4.5 &
    \textbf{DSP}: 0
    \end{tabular*}%
} \\
\hline
\end{tabularx}
\end{table}

\newpage

\subsection{Scalability to Larger Code Distances}

\begin{figure}[htpb]
    \centering
    \includegraphics[width=\columnwidth]{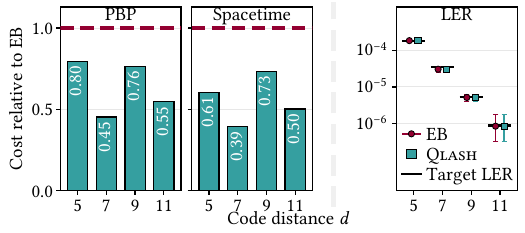}
    \vspace{-25pt}
    \caption{Costs per delivered LBP, normalized to EB at each code distance with matched delivered-LBP infidelity. Error bars indicate 95\% confidence intervals.}
    \label{fig:17}
    \vspace{-10pt}
\end{figure}

We evaluate whether \design{} retains its resource advantage as the delivered code distance increases from $5$ to $11$ and the target LBP infidelity decreases. The delivery requirements are the data-block code distance and an infidelity target of $0.1 \times LER_{\mathrm{data}}$, where $LER_{\mathrm{data}}$ denotes the data-block logical error rate.
At $p_{\mathrm{Bell}}=0.05$ and $p=0.001$, \design{} uses approximately $20$--$55\%$ fewer PBPs and $27$--$61\%$ less active spacetime per delivered LBP than EB as shown in Figure~\ref{fig:17}. At $d=11$, both point estimates fall below $10^{-6}$, while \design{} still reduces PBP consumption by approximately $45\%$ and active spacetime by $50\%$. Each comparison matches measured delivered-LBP infidelity between the methods. Both $d=7$ points lie slightly below the nominal target, and the plotted error bars cross the targets at several distances. So, cost savings persist at larger distances.

\subsection{Sensitivity to Feedback Latency}\label{sec:6.6}
The 25 ns latency in Section~\ref{sec:6.4} covers action selection after the complementary gap is available. Computing the gap requires additional MWPM decodings~\cite{sunami2025,meister2024soft}. Fixed-round EB waits only for final acceptance, while \design{} waits at intermediate decisions. We freeze policies and include idle noise and discarded attempts, assuming no QEC overlaps waits. SC and NA use assumed 1\,$\mu$s and 1\,ms cycles with a shared per-cycle noise model~\cite{google2024below,bluvstein2026architecture}. Table~\ref{tab:4} shows that at $10\,\mu$s, \design{} uses $2.26\times$ the spacetime of EB and more PBPs in SC, with lower costs in NA.
Both SC methods have matched delivered-LBP infidelity, with both estimates above the nominal target at this delay.
Accelerating the same gap can shorten waits without changing the interface, while alternative confidence signals require recalibration and recompilation~\cite{meister2024soft,xie2026postselection}.

\begin{table}[b]
\centering
\vspace{-10pt}
\caption{Preparation costs per delivered LBP under assumed
feedback delays. QEC cycles are 1\,$\mu$s for SC and 1\,ms
for NA. The unit of spacetime volume is qubit-rounds.}
\vspace{-8pt}
\label{tab:4}
\setlength{\tabcolsep}{2.5pt}
\renewcommand{\arraystretch}{1.1}
\begin{tabularx}{240pt}{
    |>{\raggedright\arraybackslash}X|c|r|r|r|r|
}
\hline
\multirow{2}{*}{    \shortstack{\textbf{Hardware}\\$p_{\mathrm{Bell}}=5\times10^{-2}$}
}
& \multirow{2}{*}{
    \shortstack{\textbf{Exposed}\\\textbf{delay} ($\mu$s)}
}
& \multicolumn{2}{c|}{\textbf{Spacetime Vol.} $\downarrow$}
& \multicolumn{2}{c|}{\textbf{\# PBPs} $\downarrow$} \\
\cline{3-6}
& & {EB\,\,\,\,\,\,} & \design{}\,\,
  & {EB\,\,\,} & \design{} \\
\hline\hline
\multirow{3}{*}{
    \shortstack[l]{Super-\\conducting (SC)}
}
& 0.1
& 5,044.5 & \textbf{1,715.9}
& 175.9 & \textbf{117.3} \\
\cline{2-6}
& 1
& 5,658.1 & \textbf{3,115.3}
& 175.5 & \textbf{121.9} \\
\cline{2-6}
& 10
& \textbf{12,381.1} & 28,005.1
& \textbf{182.4} & 206.6 \\
\hline\hline
\multirow{3}{*}{
    \shortstack[l]{Neutral\\atom (NA)}
}
& 0.1
& 4,974.9 & \textbf{1,568.8}
& 175.9 & \textbf{116.9} \\
\cline{2-6}
& 1
& 4,975.5 & \textbf{1,570.0}
& 175.9 & \textbf{116.9} \\
\cline{2-6}
& 10
& 4,981.8 & \textbf{1,583.4}
& 175.9 & \textbf{116.9} \\
\hline
\end{tabularx}
\end{table}

\subsection{Sensitivity to Tradeoff Coefficient ($w$)}
\begin{figure}[htpb]
    \centering

    \vspace{-10pt}
    \includegraphics[width=\columnwidth]
        {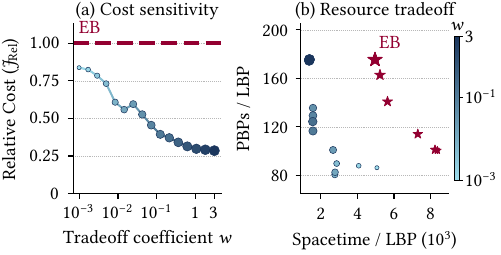}
        \vspace{-20pt}
    \caption{Impact of tradeoff coefficient $w$: (a) cost relative to EB ($J_{\text{Rel}})$;
    (b) PBPs and spacetime ($10^3$ qubit-rounds) per LBP.
    Larger markers and darker blue indicate larger $w$.}
    \label{fig:18}
    \vspace{-15pt}
\end{figure}

 We vary $w$ in Equation~\eqref{eq:5} to test whether \design{}'s cost advantage persists under different resource priorities, retuning \design{} and seven-round EB at each weight. Figure~\ref{fig:18}(a) shows that \design{} achieves lower estimated weighted cost at every tested $w$, including a $61\%$ reduction at our default $w_0=0.122$. Figure~\ref{fig:18}(b) shows the cost of prioritizing spacetime further: at high weights, PBP consumption reaches roughly EB's level for only modest additional spacetime savings over the default. We retain $w_0$ because it preserves substantial PBP savings while capturing most of the spacetime reduction achieved at high weights.

\subsection{Resource Estimation for FTQC Applications}

\noindent\textbf{Setup.} We evaluate QFT~\cite{nielsen2010quantum}, the CDKM adder (ADD)~\cite{cuccaro2004ripple}, a multiplier (MUL)~\cite{li2023qasmbench,nwqbench2021multiplier}, and Hubbard hopping (HUB)~\cite{harkness2026ftcircuitbench} using 36 logical qubits equally distributed across nine QPUs in a $3\times3$ grid. We use NWQEC~\cite{wang2025tableau} for Clifford+$T$ circuit decomposition and IRIS~\cite{yun2026iris} for DQC compilation. We compare \design{} against EB~\cite{sunami2025} under the two device settings in Table~\ref{tab:5}(a). $T$ gates consume magic states prepared by $T$ factories. For each device setting, we evaluate decentralized (DT) and centralized (CT) factory placements, as shown in Table~\ref{tab:5}(b). In DT, factories supply magic states locally within each QPU~\cite{saadatmand2026modularity}. In CT, shared factories supply the DQC~\cite{keskin2025lattice}, and we use LBPs to teleport their outputs to the target QPUs.

\vspace{0.05in} \noindent\textbf{Results.} \design{} cuts both costs relative to EB across all benchmarks under both device settings and factory placements, as shown in Table~\ref{tab:5}(c). For each benchmark under DT or CT, \design{} and EB deliver the same number of LBPs. The PBP reduction thus comes from using fewer PBPs per delivered LBP. Under DT, LBPs are needed only for remote operations on data qubits. Under CT, each $T$ gate also needs magic-state teleportation, creating extra LBP demand. For the $T$-intensive QFT benchmark, the total demand grows from $349$ LBPs under DT to $166{,}662$ LBPs under CT, because a $T$ state may pass via intermediate QPUs and uses one LBP per link. In the evaluated CT configurations, $T$ gates wait for the factory links to supply LBPs for these transfers. Using fewer PBPs per delivered LBP shortens this wait and cuts the time spent maintaining data qubits. With the parameters of \citet{stephenson2020}, QFT spacetime is about $24\%$ lower than EB under CT, compared with about $1\%$ under DT. Across the four benchmarks, CT yields geometric-mean spacetime reductions of $23.8\%$ with the parameters of \citet{stephenson2020} and $16.4\%$ with those of \citet{main2025}.

\vfill

\begin{table}[t]
\centering
\caption{(a) Trapped-ion DQC parameters. (b) How $T$-Factories are placed. (c) Estimated costs relative to EB with distance-7 LBPs at matched infidelity (target infidelity: $3.5\times10^{-5}$).}
\vspace{-10pt}
\label{tab:5}
\small
\setlength{\tabcolsep}{2.5pt}
\renewcommand{\arraystretch}{1.02}
\begin{tabularx}{240pt}{
    |>{\raggedright\arraybackslash}X|c|c|c|
}
\hline
\textbf{(a) Device Configuration}
& \shortstack{\textbf{QEC cycle}\\[-3pt]\textbf{time}}
& \shortstack{\textbf{PBP}\\[-1pt]\textbf{infidelity}}
& \shortstack{\textbf{PBP gen.}\\[-3pt]\textbf{time}} \\
\hline\hline
\citet{stephenson2020}
& \multirow{2}{*}{$2$\,ms}
& 6.0\,\%
& 5.5\,ms \\
\cline{1-1}\cline{3-4}
\citet{main2025}
&
& 2.85\,\%
& 103\,ms \\
\hline
\end{tabularx}
\\[5pt]
\begin{tabularx}{240pt}{
    |>{\raggedright\arraybackslash}X|
}
\hline
\textbf{(b) \textit{T}-Factory Placement}
($D$: data qubits, $T$: \textit{T} factories, Yellow: LBPs)\\
\hline
\includegraphics[width=\linewidth]{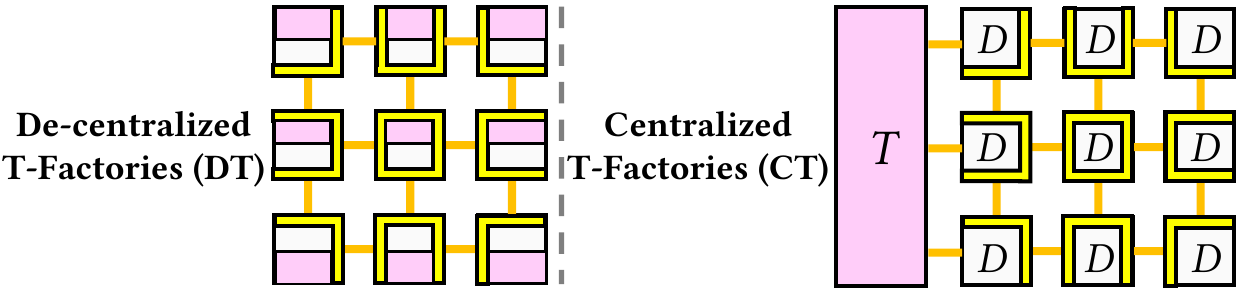} \\
\hline
\end{tabularx}
\\[5pt]
\begin{small}%
\setlength{\tabcolsep}{1.4pt}%
\resizebox{240pt}{!}{%
\begin{tabular}{
    |l|l|c|c|c|c|
}
\hline
\multicolumn{1}{|c|}{\multirow{2}{*}{\shortstack{\textbf{(c) Benchmarks}}}}
& \multirow{2}{*}{
    \shortstack{\textbf{Device}}
}
& \multicolumn{2}{c|}{\textbf{Spacetime Vol.} $\downarrow$}
& \multicolumn{2}{c|}{\textbf{\# PBPs} $\downarrow$} \\
\cline{3-6}
\multicolumn{1}{|c|}{} & & \multicolumn{1}{c|}{DT} & \multicolumn{1}{c|}{CT}
  & \multicolumn{1}{c|}{DT} & \multicolumn{1}{c|}{CT} \\
\hline\hline
\multirow{2}{*}{
    \shortstack[l]{QFT}
}
& \citet{stephenson2020}
& 0.99$\times$ & 0.76$\times$ & 0.78$\times$ & 0.78$\times$ \\
\cline{2-6}
& \citet{main2025}
& 0.96$\times$ & 0.84$\times$ & 0.85$\times$ & 0.85$\times$ \\
\hline\hline
\multirow{2}{*}{
    \shortstack[l]{ADD}
}
& \citet{stephenson2020}
& 0.93$\times$ & 0.77$\times$ & 0.77$\times$ & 0.78$\times$ \\
\cline{2-6}
& \citet{main2025}
& 0.82$\times$ & 0.83$\times$ & 0.84$\times$ & 0.84$\times$ \\
\hline\hline
\multirow{2}{*}{
    \shortstack[l]{MUL}
}
& \citet{stephenson2020}
& 0.81$\times$ & 0.76$\times$ & 0.78$\times$ & 0.78$\times$ \\
\cline{2-6}
& \citet{main2025}
& 0.84$\times$ & 0.83$\times$ & 0.85$\times$ & 0.85$\times$ \\
\hline\hline
\multirow{2}{*}{
    \shortstack[l]{HUB}
}
& \citet{stephenson2020}
& 0.99$\times$ & 0.76$\times$ & 0.78$\times$ & 0.78$\times$ \\
\cline{2-6}
& \citet{main2025}
& 0.94$\times$ & 0.84$\times$ & 0.84$\times$ & 0.85$\times$ \\
\hline\hline
\multirow{2}{*}{\shortstack[l]{\textbf{Geometric}\\\textbf{Mean}}}
& \citet{stephenson2020}
& \textbf{0.93$\times$} & \textbf{0.76$\times$} & \textbf{0.78$\times$} & \textbf{0.78$\times$} \\
\cline{2-6}
& \citet{main2025}
& \textbf{0.89$\times$} & \textbf{0.84$\times$} & \textbf{0.84$\times$} & \textbf{0.85$\times$} \\
\hline
\end{tabular}}%
\end{small}
\end{table}

\section{Complexity Analysis}

 \noindent\textbf{Online Policy Evaluation.}
After the decoder computes $g[t]$, locating its interval by binary search takes $\mathcal{O}(\log N_g)$ time, where $N_g$ counts current-gap intervals and separate values for unavailable gaps. Once the policy entry for $x[t]$ is located, reading its probabilities and sampling among at most four actions take $\mathcal{O}(1)$ time. For fixed $N_g$, policy evaluation is therefore $\mathcal{O}(1)$, independent of code distance and the number of preparation states.

\vspace{0.03in} \noindent\textbf{Complementary Gap Computation.}
Following EB~\cite{sunami2025}, each gap requires four MWPM decodings of the accumulated syndrome history $s[1:t]$. Decisions after every round therefore require $4R$ decodings over $R$ QEC rounds. Each QPU produces at most $d_f^2-1$ binary stabilizer outcomes per round, so storing both histories requires at most $2(d_f^2-1)R$ bits, or $\mathcal{O}(d_f^2R)$ memory.

\vspace{0.03in} \noindent\textbf{Offline LP Computation and Policy Storage.}
Let $N$ count preparation states and code growth plans. With at most four actions per state, the LP in Equation~\eqref{eq:5} has $\mathcal{O}(N)$ variables and constraints. A continuation action fixes the next code configuration and round counts. Each current-gap interval lies within one previous-gap interval, fixing the next previous-gap interval. Only the next current-gap interval varies, giving at most $N_g$ next states. Each plan likewise has at most $N_g$ states after its first round. The constraint matrix therefore has $\mathcal{O}(N_gN)$ nonzero coefficients. Given the sparse model and state indices, assembling it takes $\mathcal{O}(N_gN)$ operations. For rational input of total bit length $L$, a theoretical upper bound for solving the LP is $\mathcal{O}(N^{3.5}L)$ arithmetic operations~\cite{karmarkar1984}. Converting the solution into action and plan-selection probabilities requires $\mathcal{O}(N)$ arithmetic operations and $\mathcal{O}(N)$ stored values.

\newpage
\section{Related Work}\label{sec:8}

\noindent\textbf{LBP Preparation.}
Pattison et al. select distillation codes to reduce PBP consumption with a limited number of logical qubits~\cite{pattison2025}.
Grow-and-distil selects distillation and code growth in advance to increase the LBP delivery rate for a given number of physical qubits and target LBP infidelity~\cite{marqversen2025}.
Direct code projection prepares LBPs by measuring local parity checks and decoding the joint syndrome from both QPUs~\cite{bonillaataides2025,maeda2025}.
EB initializes a corner region of each code block with PBP halves and the remaining data qubits locally.
It applies complementary-gap postselection after a fixed number of QEC rounds at the target distance~\cite{sunami2025}.
Using the complementary gap after each QEC round, \design{} can discard LBPs before code growth or deliver them at the target distance without further rounds.

\vspace{0.05in}
\noindent\textbf{Adaptive QEC and State Preparation.}
Syndrome-guided control has been widely studied in QEC and state preparation, including magic-state postselection~\cite{bombin2024postselection,mehta2026practical,poster2026gappriced}.
Prior work varies the number of QEC rounds~\cite{chamberland2022building,tansuwannont2023adaptive,akahoshi2026runtime} and selects which parity checks to measure~\cite{berthusen2025adaptive}.
The complementary gap is also used to choose between MWPM and a more accurate decoder~\cite{toshio2025switching}, while a spatiotemporal variant determines whether to repeat decoding with a longer syndrome history~\cite{mishima2026adaptive}.
Chau and Ho measure additional parity checks after decoding fails and discard selected Bell pairs if decoding fails again~\cite{chau2011practical}.
Q3DE uses syndromes to detect locally increased physical error rates and temporarily grow the affected code blocks~\cite{suzuki2022q3de}, while adaptive surface codes change parity checks to exclude defective qubits~\cite{siegel2023adaptive}.
Mehta et al. propose syndrome-dependent stopping and code growth for magic-state cultivation, leaving policy optimization to future work~\cite{mehta2026practical}.
To the best of our knowledge, \design{} is the first LBP preparation method to jointly optimize the starting code and a syndrome-dependent policy for remeasurement, code growth, discard, and delivery after each QEC round.

\vspace{-5pt}
\section{Concluding Remarks}

\noindent \textbf{Summary.} We presented \design{}, a compiler pass for syndrome-adaptive LBP preparation. It jointly optimizes initial preparation plans and subsequent actions under a predicted infidelity constraint, using intermediate syndrome information to determine how much further preparation each attempt receives. For three PBP infidelities, distance-7 circuit simulations at
matched delivered-LBP infidelity show 18.8-25.2\% fewer PBPs and
19.6-68.9\% less active spacetime per delivery than EB, while \design{} takes 25\,ns to make a per-round decision once the complementary gap arrives.

\vspace{0.05in}
\noindent \textbf{Limitation and Future Work.} Although \design{} incurs a negligible latency of 25\,ns once the complementary gap is available, computing the gap after each QEC round can be a major bottleneck. The resulting idle time can increase LBP infidelity and, consequently, substantially increase LBP preparation overheads. This issue is particularly significant for superconducting systems with short QEC cycle times. Faster complementary-gap computation is therefore an important challenge. We leave it for future work. %

\vspace{-0.1in}
\section*{Acknowledgements}
This research was supported by the DARPA HARQ Program, under Agreement No. HR0011-26-9-E096.
This research was also done using services provided by the OSG Consortium~\cite{osp1,osp2,osp3,osp4}, which is supported by the National Science Foundation awards \#2030508 and \#2323298.
We appreciate CloudLab~\cite{cloudlab} for providing computing clusters and storages.

\appendix

\section{\design{} Algorithm}
Algorithm~\ref{alg:1} gives the overview of \design{}.
\vspace{-10pt}
\begin{algorithm}[htpb]
\SetAlgoNoLine
\SetAlgoNoEnd
\DontPrintSemicolon
\SetNlSty{}{}{:}
\SetFuncSty{textit}
\SetAlgoCaptionSeparator{}
\SetKwProg{Fn}{function}{}{}
\SetKwFunction{Qlash}{\design{}}
\SetKwFunction{Prepare}{Prepare\_LBP}
\SetKwFunction{Fit}{Fit\_Empirical\_Model}
\SetKwFunction{Solve}{Solve\_LP}
\SetKwFunction{Calibrate}{Calibrate\_Limit}
\SetKwFunction{Mixture}{Cheapest\_Policy\_Pair}
\SetKwFunction{Round}{QEC\_Round}
\SetKwFunction{Gap}{Complementary\_Gap}
\caption{\design{}}
\label{alg:1}
\textbf{Input:} Code growth plans ($\mathcal{I}$), Training data ($\mathcal{D}_{\mathrm{tr}}$), Calibration data ($\mathcal{D}_{\mathrm{cal}}$), Cost weight ($w$), Target infidelity ($\varepsilon$)\;
\textbf{Output:} Policy pair ($\pi_A$, $\pi_B$), Mixture probability ($q$)\;
\Fn{\Qlash{$\mathcal{I}$, $\mathcal{D}_{\mathrm{tr}}$, $\mathcal{D}_{\mathrm{cal}}$, $w$, $\varepsilon$}}{
  $(\mu,\widehat P,\widehat p_{\mathrm{fail}})\gets$ \Fit{$\mathcal{I}$, $\mathcal{D}_{\mathrm{tr}}$}\;
  $\Pi\gets\{$\Solve{$\mu,\widehat P,\widehat p_{\mathrm{fail}},w',\varepsilon'$}$:(w',\varepsilon')\in\mathcal{W}\times\mathcal{E}\}$\;
  $\varepsilon^\ast\gets$ \Calibrate{$\Pi$, $\mathcal{D}_{\mathrm{cal}}$, $w$, $\varepsilon$}\;
  \Return \Mixture{$\Pi$, $\mathcal{D}_{\mathrm{cal}}$, $w$, $\varepsilon^\ast$}\;
}
\Fn{\Prepare{$\pi_A$, $\pi_B$, $q$}}{
  \Repeat{$a=\textsf{Deliver}$}{
    $\pi\gets\pi_A$ with probability $q$, else $\pi_B$;\ $i\sim z_i/\sum_j z_j$;\ $t\gets0$\;
    \Repeat{$a\in\{\textsf{Deliver},\textsf{Discard}\}$}{
      $t\gets t+1$;\ $s[t]\gets$ \Round{$c[t]$};\ $H\gets s[1{:}t]$\;
      $g[t]\gets$ \Gap{$H$};\ form $x[t]$ by Eq.~\eqref{eq:2}\;
      $a\sim\pi(\cdot\mid x[t])$\;
    }
  }
  \Return the LBP\;
}
\end{algorithm}
\vspace{-10pt}

\section{Why Does Growth Add No PBPs?}
\begin{figure}[b]
    \centering
    \includegraphics[width=\columnwidth]{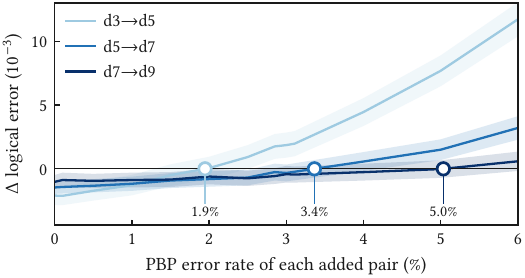}
    \vspace{-20pt}
    \caption{Change in raw LBP infidelity when PBPs fill the new data qubits at growth. The shaded bands indicate 95\% CIs. The circles mark the break-even PBP error rate.}
    \label{fig:19}
\end{figure}

\design{} does not add PBPs while growing the patch, because added PBPs lower the raw LBP infidelity only below an error rate that current PBPs exceed. We evaluate how added PBPs change the raw LBP infidelity by comparing two growth arms on the same admitted roots. When the code grows, all new data qubits are filled with PBPs in one arm and with local states in the other; growth from d3, d5, and d7 adds 16, 24, and 32 PBPs. Figure~\ref{fig:19} gives two takeaways.
First, added PBPs help only when they are nearly noiseless: the gain at 0\% vanishes above a break-even PBP error rate of 1.9\%, 3.4\%, and 5.0\% for d3$\rightarrow$d5, d5$\rightarrow$d7, and d7$\rightarrow$d9.
Second, PBP error rates are reported at 2.85\%--6\%~\cite{stephenson2020,main2025}, where adding PBPs gives no gain outside the 95\% interval and raises the raw LBP infidelity above 1.9\%, 3.4\%, and 5.0\% for d3$\rightarrow$d5, d5$\rightarrow$d7, and d7$\rightarrow$d9.

\section{Sensitivity to Number of Training Samples}

\begin{figure}[htpb]
    \vspace{-5pt}
    \centering\includegraphics[width=\columnwidth]{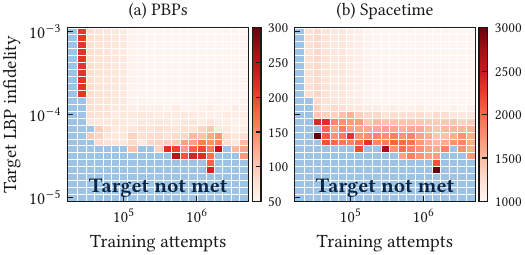}
    \vspace{-20pt}
    \caption{(a) PBPs and (b) active spacetime per delivered LBP at distance~7, by training attempts per design and target LBP infidelity. Each cell is the median over training subsets whose policy meets the target with a 95\% upper bound.}
    \label{fig:20}
\end{figure}

The number of training attempts determines how many observations support the transition and logical failure estimates in \design{}. In Figure~\ref{fig:20}, some larger training budgets yield policies that meet tighter infidelity targets. Near the boundary between targets met and missed, target attainment varies across neighboring budgets, and the costs of successful settings can differ. At looser targets, resource costs vary little across the larger training budgets.

\section{Sensitivity to Number of Gap Intervals}

\begin{figure}[htpb]
    \centering

    \vspace{-5pt}
    \includegraphics[width=\columnwidth]{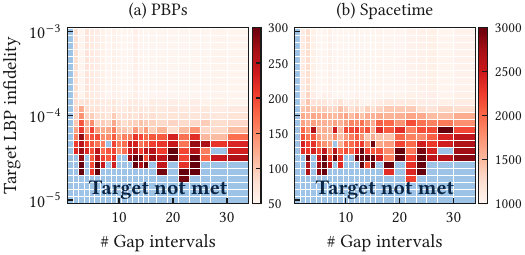}
    \vspace{-20pt}
    \caption{(a) PBPs and (b) active spacetime per delivered LBP at distance~7, by the number of equal-width intervals of the current and previous gaps and target LBP infidelity. Each cell is the median over training subsets whose policy meets the target with a 95\% upper bound.}
    \label{fig:21}
\end{figure}

Finer gap intervals let \design{} distinguish gap values that would otherwise share a preparation state, but can leave fewer training observations per state. Figure~\ref{fig:21} shows similar resource costs across many interval counts at looser targets. At stricter targets, some intermediate counts meet targets that finer partitions miss. Thus, increasing the number of intervals does not consistently improve target attainment or reduce preparation costs.

\section{Disclosure of AI Use}
While working on this paper, the authors utilized Claude 5.5 Opus and GPT 6.1 Sol for software implementation, brainstorming, and discussion of ideas to improve their understanding.

\bibliographystyle{unsrtnat}
\bibliography{refs}

\end{document}